\documentclass[ conference]{IEEEtran}
\IEEEoverridecommandlockouts
\usepackage{cite}
\usepackage{amsmath,amssymb,amsfonts}
\usepackage{algorithmic}
\usepackage{graphicx}
\usepackage{textcomp}
\usepackage{xcolor}
\usepackage{enumitem}
\usepackage{threeparttable}
\usepackage{tabularx}
\usepackage{url}
\usepackage{balance}
\usepackage{booktabs}
\def\BibTeX{{\rm B\kern-.05em{\sc i\kern-.025em b}\kern-.08em
    T\kern-.1667em\lower.7ex\hbox{E}\kern-.125emX}}

\newcommand{\ie}{\textit{i.e.,} }

\newcommand{\new}[1]{{ }}

\begin{document}

\title{Exploring the Impact of Integrating UI Testing in
CI/CD Workflows on GitHub
% {\footnotesize \textsuperscript{*}Note: Sub-titles are not captured for https://ieeexplore.ieee.org  and
% should not be used}
% \thanks{Identify applicable funding agency here. If none, delete this.}
}
\IEEEpeerreviewmaketitle
\author{\IEEEauthorblockN{Xiaoxiao Gan}
\IEEEauthorblockA{\textit{Virginia Tech} \\
}
\and
\IEEEauthorblockN{Chris Brown}
\IEEEauthorblockA{\textit{Virginia Tech} \\
}

}

\maketitle

\begin{abstract}
Background: User interface (UI) testing, which is used to verify the behavior of interactive elements in applications, plays an important role in software development and quality assurance. However, little is known about the adoption of UI testing frameworks in continuous integration and continuous delivery (CI/CD) workflows and their impact on open-source software development processes.

Objective: We aim to investigate the current usage of popular UI testing frameworks–Selenium, Playwright and Cypress–in CI/CD pipelines among GitHub repositories. Our goal is to understand how UI testing tools are used in CI/CD processes and assess their potential impacts on open-source development activity and CI/CD workflows.

Method: We propose an empirical study to examine GitHub repositories that incorporate UI testing in CI/CD workflows. Our exploratory evaluation will collect repositories that implement UI testing frameworks in configuration files for GitHub Actions workflows to inspect UI testing-related and non-UI testing-related workflows. Moreover, we further plan to collect metrics related to repository development activity and GitHub Actions workflows to conduct comparative and time series analyses exploring whether UI testing integration and usage within CI/CD processes has an impact on open-source development. 
\end{abstract}

\begin{IEEEkeywords}
CI/CD, UI Testing, Software Engineering
\end{IEEEkeywords}

\section{Introduction} 
% / 1page
% RQ2: How does UI testing usage in GitHub Actions affect development processes? \\
% number of commits, 
% number of pull requests, 
% successful rate of pull requests, 
% number of merged/rejected pull requests
% Number of Created Issues relevant to UI testing

% RQ3: How does UI Testings usage in GitHub Actions affect CI/CD workflows?
% Test execution time after ui testing
% number of ui tests among all the tests 
% overall trends of ui tests???? 

Continuous Integration and Continuous Delivery (CI/CD) is essential in modern software development~\cite{7884954}, especially within open-source software~\cite{hilton2016usage}. A typical CI/CD pipeline consists of several phases to stage, build, and test code changes before deploying to users \cite{rahman2015synthesizing}. CI/CD enables more frequent and reliable software releases and reduces the cost of development. For example, a case study at the HP LaserJet Firmware division~\cite{Humber16} reports adopting CI/CD practices reduced overall development costs by 40\% and increased programs under development by 140\%. CI/CD has also been shown to improve productivity and quality for open-source projects~\cite{vasilescu2015ci,fairbanks2023analyzing}. 

Automated testing in CI/CD pipelines is critical for development and business success~\cite{headspin}. For instance, the Volkswagen IT Cloud group found testing in CI/CD prevents bottlenecks and provides faster product delivery~\cite{poth2018deliver}. One specific type of testing is user interface (UI) testing. UIs are the primary way humans interact with software, which handle the input and output for programs~\cite{idf}. UI testing plays a pivotal role in assessing the user interface of open-source applications to ensure the features of applications that are visible to users (\ie buttons, text boxes, images, etc.) are functional and meet business requirements~\cite{browserstack}. UI testing has been shown to be useful for verifying user interfaces in software, particularly when integrated in CI/CD pipelines~\cite{Millian22,Brandon22}. %For instance, a case study \cite{6993503} examined introducing continuous integration and UI automated testing approaches in the software development process for a complicated smart grid support system (D5000 systems), finding a significant increase in development team productivity and code quality, with the reduction of the risk in project execution.

However, UI testing is challenging in software development. Prior work shows UI testing conflicts with regression testing--a type of functional software testing commonly used in CI/CD pipelines~\cite{parnin2017top}--due to its dependence on user input, output relying on the layout of elements, and rapid changes made to UIs by developers~\cite{memon2002gui}. Developers also report writing UI tests is difficult in CI/CD tools, such as TravisCI~\cite{widder2019pain}. To avoid these challenges, practitioners often employ manual functional testing methods for user interfaces, expressing concerns and grievances about the limitations inherent to automated UI testing~\cite{10.1145/3474624.3474640}. For instance, Elazhary et al. report developers find the efforts needed to integrate UI testing into CI/CD workflows are not worth the benefits~\cite{elazhary2021uncovering}. This lack of confidence and support for automated UI testing can prevent practitioners from testing UIs in their CI/CD workflows, ultimately impacting the quality~\cite{banerjee2013graphical} and security~\cite{luo2017hindsight} of software products. 

Prior work suggests UI testing frameworks, such as Selenium, are widely used in open-source software~\cite{garcia2020survey}. \textbf{However, the usage and impact of UI testing in CI/CD pipelines within open-source repositories remains a subject of investigation.} Sources suggest most open-source software is developed for commercial use~\cite{octoverse2022}, with 96\% of proprietary projects containing open-source code~\cite{synopsys2023report}. CI/CD adoption is also increasing rapidly in open-source projects. For example, research characterizing CI/CD usage in open-source software shows the adoption of CI/CD is increasing, with many projects using multiple CI/CD technologies simultaneously~\cite{da2024chronicles}.

To understand whether UI testing in CI/CD workflows impacts open-source development, we aim to conduct a comprehensive investigation into the utilization and impact of UI testing frameworks within configuration files for CI/CD pipelines. We will devise an exploratory study to collect and analyze data from GitHub projects using popular UI testing frameworks---namely Selenium, Playwright, and Cypress---in CI/CD configurations with GitHub Actions. GitHub is the predominant platform for open-source projects~\cite{octoverse2022}, and GitHub Actions provide functionality for developers to automate CI/CD tasks within repositories~\cite{kinsman2021software}. Through these data, we will investigate whether the integration of UI testing in CI/CD impacts development activity and CI/CD workflows. Our results will provide empirical insights on the effects of UI testing in CI/CD workflows and motivate solutions to enhance the assessment of UIs in modern software development.

%compiled a dataset of GitHub projects which use Selenium-based technologies and popular CI/CD platforms. For RQ1, we collected and analyzed configuration scripts for three CI/CD platforms--GitHub Actions, Travis CI, and Jenkins--that incorporate Selenium to systematically identify and categorize usage patterns. The results indicate common patterns of how Selenium is used in open-source repositories, such as executing commands in CI/CD environments. We further analyzed repositories that incorporate Selenium into GitHub Actions configuration scripts to explore the impact of UI testing on development processes and CI/CD workflows for RQ2 and RQ3. We found while there significant impact on some development activities such as commit frequency or he frequency of closed pull requests or merged pull requests on repositories, it shows minimal influence on the creation of new issues or pull requests. Additionally, we noticed a negative impact on workflow run time, although this did not affect the final outcomes. Our main contributions can be described as follows:

% \begin{enumerate}
%     \item An empirical study of Selenium usage in CI/CD platforms, offering nuanced insights into its diverse applications.
%     \item An understanding of the impact of incorporating Selenium testing into CI/CD workflows on development activity and project builds.
%     \item Implications for developers considering integrating UI testing into CI/CD processes and mechanisms to support them.
% \end{enumerate}

\section{Background and Related Work}
\begin{figure*}
    \centering
    \includegraphics[width=0.75\linewidth]{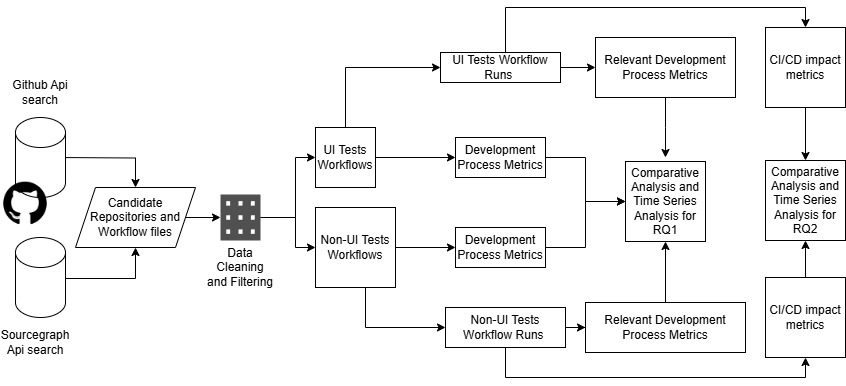}
    \caption{Study Design Overview}
    \label{fig:enter-label}
\end{figure*}
\subsection{ UI Testing Frameworks}

User interface testing frameworks are browser automation tools that provide functionality to automatically assess the behavior of UIs in programs. Our proposed study will focus on three web-based UI testing platforms: Selenium, Cypress, and Playwright. We select these frameworks based on their high popularity and usage in practice, denoted in grey literature as the current most popular UI testing tools~\cite{tools,axelerant}.

\subsubsection{Selenium}

Selenium\footnote{\url{https://www.selenium.dev/}} is an open-source framework for automating browsers to support testing UI elements of programs~\cite{vila2017automation} and other web automation tasks~\cite{seleniumboring}. Prior work has extensively explored Selenium in a variety of software testing and development contexts (\ie ~\cite{shariff2019improving},~\cite{garcia2022Selenium},~\cite{presler2019wait}). For instance, prior work explores challenges and solutions for end-to-end testing---a type of testing for verifying the complete functionality of software systems---with Selenium~\cite{leotta2023challenges}. Research also demonstrates the value of Selenium for UI testing in Agile software development processes~\cite{holmes2006automating}. 

\subsubsection{Cypress} Cypress\footnote{\url{https://www.cypress.io/}} is an end-to-end JavaScript testing tool that supports automating, debugging, and testing browser features of software. Cypress claims to be able to provide ``fast, easy and reliable testing for anything that runs in a browser''. Researchers have investigated the capabilities of Cypress in various UI testing scenarios~\cite{taky2021automated,cao2022using}. For instance, Pelivani and colleagues compared the performance of Cypress for UI testing against state-of-the-art systems, such as Selenium~\cite{pelivani2022comparative} 

\subsubsection{Playwright} Playwright\footnote{\url{https://playwright.dev/}} is an open-source platform introduced by Microsoft. This system provides functionality to support web testing and automation across browsers and programming languages. There is limited research exploring this platform in empirical and research settings. However, despite the relatively recent release of Playwright in 2020, it already has surpassed more established frameworks in adoption~\cite{playwright-adoption} and browser automation features~\cite{garcia2024browser}. Practitioner-focused sources suggest Playwright is faster and more usable that existing UI testing platforms,  providing better support for verifying the behavior of user interfaces in modern software development processes~\cite{axelerant,testomatio}. 

\subsection{ CI/CD Platforms }

CI/CD platforms are services and tools that provide support for automating various activities related to continuous integration software projects~\cite{singh2019comparison}, including testing. For this research, we explore the usage of UI testing in the GitHub Actions CI/CD platform. Prior work suggests GitHub Actions is one of the most popular services used to support CI/CD tasks in open-source software~\cite{golzadeh2022rise,wessel2022github}. While the nature of our work is exploratory, we acknowledge focusing only on three UI testing frameworks and one CI/CD platform limits the generalizability of our findings.

\subsubsection{GitHub Actions}

GitHub Actions\footnote{\url{https://github.com/features/actions}} 
 is an automated workflow and CI/CD platform provided by GitHub, allowing developers to automate various tasks, build, test, and deploy code directly within their GitHub repositories. GitHub Actions has been the subject of interest in numerous studies given its impact on software development and deployment. Recent work, such as that by Kinsman et al.~\cite{kinsman2021software}, investigates developer's behavior and usage of GitHub Actions. Prior work has also explored the impact of GitHub Actions on pull requests--code contributions to open-source repositories--finding increased rejections and communications on contributions~\cite{wessel2022github}. In GitHub Actions, automation tasks are primarily implemented through \textit{workflows},\footnote{\url{https://docs.github.com/en/actions/writing-workflows/about-workflows}} which are triggered by repository events to automate specific tasks in \textit{workflow runs}. In this work, we aim to investigate whether the integration and usage of UI testing in CI/CD pipelines impacts open-source projects by examining UI testing-related and non-UI testing-related GitHub Action workflow runs in project repositories.

\subsection{Related Work}

Researchers have explored the impact of testing in CI/CD workflows. For instance, Beller et al. investigated the impact of the testing phase on Travis CI workflows~\cite{beller2017oops}. Their research demonstrated that the incorporation of testing resulted in a 10\% increase in the detection of failures at build time within the Travis CI environment. Moreover, Pratama et al. investigated the performance testing in CI/CD~\cite{pratama2021implementation}. Their work aimed to integrate, automate, and periodically execute test processes, but failed to yield positive results due to multiple constraining factors. This study underscores the complexities and challenges inherent in integrating CI/CD into performance testing and highlights the need for further exploration and resolution of the identified constraints to advance the field. Our research aims to extend this work by investigating the impact of integrating user interface testing, through frameworks such as Selenium, into the configuration of CI/CD workflows.

There is limited research exploring the intersection of CI/CD and UI testing. Lu et al. found automated UI testing techniques and continuous integration can improve development productivity, quality, and integration~\cite{6993503}. Prior work has further investigated testing in CI/CD workflows. Rahman et al. outlined the types of testing approaches adopted by CI/CD projects~\cite{rahman2015synthesizing}--noting the prevalence of practices such as unit testing and integration testing techniques, but a lack of UI testing. Prior work also suggests specific types of testing, such as user interfaces and accessibility, are often overlooked in large open and closed source companies that adopt rapid release processes, such as CI/CD~\cite{mantyla2015rapid}. The most closely related work is a case study investigating the integration of automated graphical UI testing in CI/CD for the D5000 smart grid support system, demonstrating an increase in productivity and code quality, while reducing risk in project execution~\cite{6993503}. However, to our knowledge their is no prior work exploring the impact of UI testing on CI/CD in open-source software.

\section{Study Design}
Our study aims to explore whether UI testing integration and usage has any impacts on open-source development activity and CI/CD pipelines, in order to identify areas to better support and streamline UI testing in modern software development. To address this goal, we present a detailed study plan to answer the proposed research questions: 

\begin{itemize}
    
\item[] \textbf{RQ1:} How does UI testing integration and usage in GitHub Actions affect development processes?
    % \new{
    %     \begin{itemize}
    %         \item[] \textbf{RQ1.1:} Compared with other non-UI testing workflows, does UI testing workflow have significant differences of developer activities in terms of number of pull request / commits / issues?
    %         \item[] \textbf{RQ1.2:} How are pull requests/ commits / issues related to UI testing?
    %         \item[] \textbf{RQ1.3:} What are the major categories, causes and strategies related to UI test-relevant pull requests / commits / issues?
    %         \item[] \textbf{RQ1.4:} Compared with other issues, are these issues takes longer to resolve?
    %     \end{itemize}
    % }
\item[] \textbf{RQ2:} How does UI testing integration and usage in GitHub Actions affect CI/CD workflows?
    % \new{
    %     \begin{itemize}
    %         \item[] \textbf{RQ2.1:} Compared with other non-UI testing workflows, does the UI test-relevant workflow show significant differences in the number of workflow runs?
    %         \item[] \textbf{RQ2.2:} Compared with other non-UI testing workflows, does the UI test-relevant workflow show significant differences in the percentage of workflow runs failure rate?
    %         \item[] \textbf{RQ2.3:} Compared with other non-UI testing workflows, does the UI test-relevant workflow show significant differences in the duration of follow-up pass execution? 
    %         \item[] \textbf{RQ2.4:} Compared with other non-UI testing workflows, does the UI test-relevant workflow show significant differences in the execution duration?
    %         \item[] \textbf{RQ1.5:} Compared with other issues, are these issues takes longer to resolve?
    %     \end{itemize}
    % }
\end{itemize}

% \todo{why these research questions?} 
An overview of our proposed methodology is presented in Figure~\ref{fig:enter-label}. This study is designed to augment our preliminary findings investigating open-source software developers' experiences and perceptions of integrating UI testing in CI/CD workflows through a survey and interview study.

\subsection{Repository Selection}
To answer our research questions, we need to collect repositories that integrate UI testing within CI/CD pipelines. We will use the GitHub REST API\footnote{\url{https://docs.github.com/en/rest?apiVersion=2022-11-28}} to collect relevant open-source repositories from GitHub. Specifically, we will leverage the API code search functionality to search UI testing-related keywords in projects' GitHub Action configuration files. To collect configuration files, we will search for repositories containing files that match the path \texttt{.github.workflows/.*.yml} or \texttt{.github/workflows/.*.yaml}. Next, we will inspect configuration files to detect whether or not they include a relevant keyword---such as  ``\texttt{selenium}'', ``\texttt{cypress}'', or \texttt{playwright}''---indicating a UI testing framework is used in the CI/CD pipeline. We chose these keywords as they are the most common and straightforward method to determine if UI testing frameworks are integrated in project configuration files for corresponding tests. For instance, if a CI/CD pipeline contains Cypress tests, it may have a command in a workflow such as `\texttt{npm install cypress}` to set up the environment for Cypress tests. We acknowledge the limitation of this approach, however we observed a lack of methods to detect whether UI tests are involved in CI/CD pipelines. We will conduct initial manual checks on a subset of repositories to verify the accuracy of our keyword-based approach in detecting projects with UI testing in CI/CD workflows---potentially pivoting to explore other options based on the preliminary results.

Since the GitHub API code search functionality only returns the first 1,000 records and does not support filtering result by date, we also plan use Sourcegraph\footnote{\url{https://sourcegraph.com/search}} to supplement our efforts to collect additional repositories with UI testing frameworks integrated in GitHub Actions workflows. Sourcegraph is a code intelligence platform that provides support for code search in open-source repositories on social coding platforms, including GitHub and GitLab \cite{Beyang2020}. This process will result in a list of candidate repositories that integrate UI testing in GitHub Action workflows, along with the corresponding configuration files.

After obtaining a list of candidate repositories, we will filter out projects to use for our evaluation using following criteria:
\begin{enumerate}
    \item The repository must have an open-source license
    \item The repository must have at least ten stars
    \item The repository must have at least five contributors
    \item The repository must not be a forked project
    \item The repository must be active in the last six months
    \item The repository must have at least five workflow runs for identified workflow configuration file
    \item The repository must have corresponding UI test files explicitly specified for execution in the workflow files.
    \item The repository must have at least one non-UI tests workflow configuration files
\end{enumerate}

Based on these criteria, we aim to collect a diverse sample of open-source repositories, consisting of projects with varying sizes, programming languages, and software domains, that utilize UI testing in CI/CD workflows. We plan to develop Python script to automate the filtering of repositories based on the first six criteria, as these are straightforward to implement programmatically. Table \ref{tab:api_table} shows potential GitHub API query options based on our filtering criteria. After the automated filtering process, two researchers will manually review the remaining repositories to assess if the repository match the seventh and eighth requirements. These criteria were determined based on similar criteria used in prior repository mining studies. We aim to select popular and actively maintained repositories on GitHub and avoid inactive and personal projects~\cite{kalliamvakou2014promises}. In necessary, we will review the resulting repositories and iteratively improve our filtering criteria to enhance the sample of projects for our evaluation.

\begin{table}[t]
    \centering
        \caption{Potential Github Search API Query Options}
    \begin{tabular}{|c|c|}
     \hline
     \textbf{Specification} &  \textbf{API Query Option} \\
     \hline
     Licenses & \textbf{license:mit license:apache-2.0 license:cc license:0bsd} \\
     \hline
     Stars & \textbf{stars:$>=$10} \\
     \hline
     Commits & \textbf{pushed:$>=$[within six months of data collection]} \\
     \hline
     Templates & \textbf{template:false} \\
     \hline
     Archived & \textbf{archived:false} \\
     \hline
    \end{tabular}
    \label{tab:api_table}
    \begin{tablenotes}
    \item * These filtering criteria and GitHub API query options are only for an example and subject to change for the actual evaluation. The GitHub REST API currently does not provide the ability to query by the number of contributors and workflow runs. However, this data can be obtained by retrieving the lists of the items and determining the size is greater than the filtered amount.
    \end{tablenotes}
\end{table}

\begin{table*}
    \centering
    \caption{RQ1: Development Activity Metrics}
    \begin{tabular}{lll}
        \toprule
        \textbf{Metric} & \textbf{Type} & \textbf{Description} \\
        \midrule
        Configuration File & string & File path(s) of the GitHub Actions configuration files \\
        Introducing Commit & string & Revision hash of the commit introducing UI testing in the configuration file \\
        Introducing Date & datetime & Date of the commit when UI testing was introduced into the GitHub Actions workflow \\
        Commits &  integer & Number of total commits, frequency, percentage of general commits and UI tests-related commits \\
        Pull Requests &  integer & Number of created pull requests  and UI tests-related pull requests  \\
        Pull Request Status & string   &  Current state of the pull request (open, closed, or merged) \\
        Created Issues & integer & Number of issues created \\
        Issue Status & string & Current state of the issue (open or closed) \\
        Issue Tags & list of strings & Tags designated for issues \\
        Closed Issues &  integer & Number of closed issues   \\
        \textcolor{black}{
        Pull Request Duration} & datetime & Duration of pull request resolution, derived by the pull request open time and closed time \\ 
        \textcolor{black}{Issue Duration} & datetime & Duration of issue resolution, derived by the issue open time and closed time \\

        \bottomrule
    \end{tabular}
    
    \label{tab:developer_metric}
\end{table*}

\begin{table*}
    \centering
    \caption{RQ2: CI/CD Workflow Metrics}
    \begin{tabular}{lll}
    \toprule
        \textbf{Metric} & \textbf{Type} & \textbf{Description}       \\
        \midrule
        Workflow ID & string & Identifier for the workflow from GitHub Actions \\
        Workflow runs & integer   & Number of workflow runs initiated \\
        % Relevant commits & integer & \todo{how to determine relevant commits} \\
        Execution Time   &  integer & Amount of time for the workflow to complete in seconds   \\
        Status &  boolean  & Outcome of the GitHub Action workflow run (pass or fail) \\
        Workflow issues & list of strings & List of issue identifiers for issues related to workflow runs \\
         \textcolor{black}{Failure Workflow Run Duration} & datetime & Duration of failure workflow run and next successful workflow run \\
        \bottomrule
    \end{tabular}
    \label{tab:CI/CD impact metric}

\end{table*}
\subsection{Data Mining and Preparation}
After identifying and filtering the candidate repositories, we will leverage the GitHub API to gather other CI/CD configuration file paths in the repositories for other GitHub Actions workflows. We will manually categorize all of the project workflows into \textit{UI test-related} and \textit{non-UI test-related} workflows. For RQ1, we will collect metrics related to developer activities, such as the number of commits, number of pull requests, number of issues, etc. These metrics will be collected across two groups---those associated with the identified UI test-related workflows or associated with other non-UI testing workflows for our data analysis. The details for the initial list of potential development activity metrics we plan to collect can be viewed in Table \ref{tab:developer_metric}.

\new{
There could be scenarios where commits may impact both UI-related issues and non-UI related issues. To address this, we plan to categorize such commit into a new category - Mixed category. Based on the number of commits of this category, we will decide if we can simply ignore this type of commit (if they are few) or if we should take them into consideration in both categories when necessary.
}

For RQ2, we will use the GitHub API to collect relevant CI/CD data from GitHub Actions, including the number of workflows runs, failure rate of workflow runs, total and average execution time of workflow runs, and relevant issues associated with the workflows file. Furthermore, these metrics will also be gathered across UI test-related and non-UI test-related workflows for comparison in the analysis phase. The details of the initial set of metrics for determining CI/CD impact can be viewed in Table \ref{tab:CI/CD impact metric}.

We will analyze each repository to determine the starting date of the UI testing in GitHub Action. To do this, we will leverage the GitHub API to pinpoint the first commit that introduced a UI testing keyword into a configuration file for GitHub Actions. Additionally, all of the metrics for development activity (RQ1) and CI/CD workflow (RQ2) will be collected both in aggregate to compare across UI testing and non-UI testing workflows in addition to being observed over time. We will conduct initial pilot tests based on the sampled repositories to determine the data frequency for observing the collected metrics before and after the introduction of UI testing in a GitHub Actions workflow file.

\subsection{Analysis and Evaluation Plan}

Our goal is to uncover patterns from the collected data to determine whether UI testing has an impact on CI/CD workflows in open-source development. We aim to use multiple analysis techniques to inspect the collected data and observe the effects of UI testing integration and usage.

\subsubsection{Comparative Analysis}

We plan to conduct comparative quantitative analyses to explore the impacts of UI testing in CI/CD on development activity and workflow runs.

\paragraph{Descriptive Statistics} First, we plan to provide an overview of all relevant metrics by calculating descriptive statistics (\ie mean and median) for both development activity-relevant and CI/CD-relevant metrics presented in Tables \ref{tab:developer_metric} and \ref{tab:CI/CD impact metric}. For example, we plan to calculate the proportion of pull requests opened related to UI testing-related workflows compared to non-UI testing-related ones for RQ1 and measure the workflow failure rate for UI testing and non-UI testing workflow failure for RQ2.

\paragraph{Inferential Statistics} To further determine whether UI tests in CI/CD pipelines effects open-source development, we plan to use inferential statistics based on our sample repositories and collected metrics to estimate characteristics of projects incorporating UI testing in CI/CD processes. We will conduct a comparative analysis will be conducted by using paired t-test or Wilcoxon Signed-Rank test to determine statistical significance ($p < 0.05$). We will conduct initial statistical tests, such as Shapiro-Wilk test, to determine the normality of the data before selecting and performing the statistical analysis. Particularly, we will observe whether UI test-relevant workflows differ significantly from non-UI test-relevant workflows in terms of the associated metrics for development activity (RQ1) and CI/CD workflows (RQ2). For RQ1, we also will observe whether UI test-related workflow runs differ in relevant commits, pull requests, and issues compared to non-UI tests workflow runs, aiming to identify whether the integration impacts the development process. Since there is no direct way to fetch commits, pull requests or issues influenced by a specific workflow run, we will identify the commit or pull request that triggers the workflow run, and filter the commits, pull requests and issues by the timestamp and analyze the commit, issue and pull requests message to assess the relevance to the workflow run.

To address RQ2, we will quantitatively observe whether UI test-related workflow run results differ with regard to execution time and failure rate compared to the non-UI test-related workflow runs by using a similar comparative analysis.

%After relevant data extraction, we will use Chi-Square statistic to measure the difference between general metrics of repository and the UI-test relevant metrics of repository \todo{This looks more UI testing impact not UI testing in CI??} 

% \subsubsection{Correlation Analysis}
% To gain deep understanding for RQ1 and RQ2, we also 
\subsubsection{Time Series Analysis}
To further explore whether there is a causal impact of changes in development activity and workflow runs regarding UI testing in CI/CD workflows, we plan to augment our evaluation by conducting time series analysis~\cite{mcdowall2019interrupted}. This will allow us to further explore if an intervention--such as the introduction of UI testing in CI/CD pipelines--has effects on the observed metrics over a period of time before and after the intervention occurred. To calculate this, we will model development and workflow metrics before and after UI testing keywords were introduced to configuration files for GitHub Actions workflows. We will conduct initial pilot studies using a random sample of repositories for our evaluation to form the time period for our analysis and better understand whether changes in development outcomes changed after UI testing integration in workflows. We will also conduct additional checks to enhance the reliability of time series findings, such as stationarity, autocorrelation, and seasonality~\cite{walls1987time}. 

\new{
We understand there is a possibility that not every repository will have enough data before and after adopting UI testing (\ie, repositories that integrated UI tests since the beginning of the project or those that integrated UI tests very recently). Therefore, we plan to select high quality repositories  that meet the requirements for this specific analysis. 

% If we still do not have enough data, our backup plan is to use Difference-in-Differences analysis to compare changes in different groups: 1) a group of projects using UI testing in CI/CD; and 2) another group of similar projects that do not integrate UI tests in CI/CD.

}

To conduct the time series analysis, we will clone and download the selected repositories to gather the commit history of CI/CD configuration files and identify the specific commit that introduced the execution of UI tests. The timestamp of this specific commit will serve as the intervention marker. We will check to ensure the repositories have a minimum number of data points to determine statistical significance with time series analyses~\cite{warner1998spectral}. Then, we will compare the key metrics with sufficient data for RQ1 (\ie number of commits, pull requests, closed issues/pull requests, etc.) and RQ2 (\ie build times, failure rate, etc.) before and after the implementation using the interrupted time series model. 

\subsubsection{Tools}
We aim to develop Python scripts leveraging the GitHub API to select relevant repositories and collect the necessary data. For the data processing and analysis, we plan to use matplotlib\footnote{\url{https://matplotlib.org/}} and pandas\footnote{\url{https://pandas.pydata.org/}} to conduct any statistics calculations and generate visualizations.

\subsection{Data Availability}

All collected and generated data and data collection scripts will be shared in a public repository and made available for long-term archiving to promote the transparency, replicability, and future extensions of this work.

\section{Threats to Validity}

There are several threats to the validity of our proposed experiment design, which we aim to mitigate in our study and address in future work. The scope of our project aims to observe whether incorporating UI tests in CI/CD pipelines impacts development activity and CI/CD processes for open-source projects. We will investigate this through an exploratory study by examining the integration and usage of UI testing frameworks, particularly Selenium, Cypress, and Playwright, in GitHub Action workflows.

\subsection{Construct Validity}
The primary construct that poses a threat to the study design is the selection of keywords used to identify the candidate repositories and differentiate between UI test-related and non-UI test-related workflows. We restrict the GitHub API code search to detect configuration files containing the ``\texttt{selenium/cypress/playwright}" keywords based on the assumption that workflow files containing such keywords are likely to use UI testing in  the CI/CD pipelines for their repositories. However, this approach has limitations---such as these keywords may only appear in the comments of workflow files or be used for different purposes, such web scraping and other web automation tasks not related to UI testing~\cite{seleniumboring}, which introduces potential false positives. To mitigate this, we plan to conduct manual reviews on both workflow files and validate the corresponding existence of UI tests files in the repositories during the filtering process after getting the initially search result. Throughout the data collection, we also will continue to explore other methods to automatically determine CI/CD configuration files are related to UI testing.

% \todo{keywords to identify candidate repositories}
\subsection{Internal Validity} This work will suffer from limitations related to mining open-source software repositories on GitHub~\cite{kalliamvakou2014promises}. We apply filtering criteria
to our initial collection of candidate repositories---based on popular open-source
licenses (\ie MIT, Apache 2.0, CC, and 0bsd), stars ($>=$ 10), contributors ($>=$ 5), workflow runs ($>=$ 5), non-forked projects, and commit dates within six months of data collection---to select popular and active GitHub
projects for our evaluation. We also collect projects with UI at least one UI test-related and one non-UI test-related workflow for our analysis. However, these filters may limit generalizability and incorporate bias in the collected data.

\subsection{External Validity} Our proposed study only focuses on open-source projects on GitHub, and our findings may not generalize to all types of software. To mitigate this, we aim to collect a diverse sample of repositories across programming languages, sizes, domains, etc. However, future work is needed to investigate the effects of UI testing on closed-source proprietary software or open-source projects hosted on other platforms (\ie GitLab\footnote{\url{https://about.gitlab.com/}}). In addition, we focus on limited UI testing platforms. We selected Selenium, Cypress, and Playwright due to their popularity in practice and on GitHub---however, our results may not generalize to other UI testing tools (\ie Puppeteer\footnote{\url{https://pptr.dev/}}). This work also only explores UI testing in one CI/CD platform---GitHub Actions. We leverage this platform due to its popularity and availability of data to explore the impact of UI testing on CI/CD pipelines. Prior work has similarly focused on examining CI/CD-related phenomena on open-source GitHub projects with a single platform (\ie~\cite{kinsman2021software, widder2018breakup}). Future work is needed to investigate the impact of UI testing in other CI/CD platforms, such as CircleCI.\footnote{\url{https://circleci.com/}}

\new{
\section{Potential Impact}
Our study will mine data from GitHub repositories to collect empirical insights on the effects of integrating UI testing frameworks in to CI/CD pipelines---exploring the integration of Selenium, Cypress, and Playwright in GitHub Actions on open-source repositories for web applications. This study will build on prior work observing the effects of CI/CD in software projects~\cite{Humber16}. We anticipate our results will have implications for development communities. For instance, outlining strategies and practices used for UI test integration in CI/CD pipelines (\ie through analysis of UI and non-UI commits) and quantifying the impact of UI testing in CI/CD workflows. We will use our experimental outcomes to provide implications and motivate guidelines for practitioners considering incorporating UI tests in CI/CD workflows.}

\new{
Our findings will also lay the groundwork for future solutions to support UI testing in CI/CD---motivating novel infrastructure to mitigate observed negative effects and challenges. Examples of possible solutions and research directions that could be implemented based on our potential findings include:
}

\new{
\paragraph{Prioritization} If we observe significant increases in CI/CD workflow metrics for projects incorporating UI tests in pipelines (\ie workflow runs and execution time), future work can explore approaches to mitigate delays in builds related to testing. For instance, prior work shows test case prioritization to reduce testing time~\cite{elbaum2002test}, including for UI-related tests~\cite{yu2019terminator,magalhaes2021ui}. To enhance CI/CD processes, studies also show prioritizing project builds can reduce time in development workflows~\cite{jin2022builds}. Recently, research shows hybrid approaches for predicting build outcomes and test outcomes can reduce project build time and costs~\cite{jin2023hybridcisave}. Our results could motivate similar approaches leveraging prioritization to reduce build and testing time in UI testing contexts.
}

\new{
\paragraph{Test Case Generation} If we observe significant increases in development activity metrics (\ie pull requests and commits), this may motivate the need to investigate automated test case generation for UI tests in CI/CD pipelines. For example, recent work explores leveraging generative artificial intelligence and large language models (LLMs) to generate test cases for code~\cite{ouedraogo2024large}---demonstrating their ability to reduce developer effort~\cite{gunnarsson2024test} and detect bugs~\cite{liu2024llm}. Similar systems can be used to automatically generate tests for UI elements of software. However, LLMs have been shown to struggle in UI development contexts~\cite{huq2023s}. Techniques such as fine-tuning or training models in specific UI testing projects~\cite{wen2024autodroid}---focusing on common UI testing frameworks such as Selenium, Cypress, and Playwright---can potentially enhance their capabilities to support UI test case generation, facilitating their integration in CI/CD pipelines.
}
\section{Conclusion}
User interfaces are the primary way that users interact with software. However, verifying the behavior of UIs is challenging and difficult to automate---becoming more complicated as the adoption of CI/CD processes rises to meet the rapid demands of software. To this end, we plan to conduct an exploratory study to investigate UI testing usage and integration within CI/CD workflows in the context of open-source repositories. Our approach includes leveraging the GitHub API to identify repositories that are representative of projects integrating UI testing in CI/CD pipelines---through adopting popular web automation and UI testing frameworks such as Selenium, Cypress, and Playwright in GitHub Actions---to mine relevant data and conduct multifaceted analyses to provide valuable insights to determine whether UI testing in CI/CD impacts development activity and CI/CD processes. Our findings will offer a deeper understanding of the role UI testing plays in the CI/CD processes and motivate solutions to enhance UI testing in modern open-source environments.

\balance

\bibliographystyle{ieeetr}
\bibliography{ref}

\end{document}